\documentclass[11pt]{article}
\usepackage{fullpage}
\usepackage{graphicx}
\usepackage{fancyhdr}
\usepackage{aas_macros}

\newcommand{\msun}{\mbox{M$_\odot$}}

\title{{\sc THE CURRENT STATE OF CLUSTER FORMATION SIMULATIONS}}
\author{\normalsize Keynote talk at the Sexten Center for Astrophysics Workshop\\
\normalsize `The Formation and Early Evolution of Stellar Clusters'\\
\normalsize 23--27 July 2012, Sexten, Italy\\\\
\large J.~M.~Diederik Kruijssen\\
\small Max-Planck Institut f\"ur Astrophysik, Garching, Germany\\
\small kruijssen@mpa-garching.mpg.de}
\date{}
\begin{document}
\maketitle

\begin{abstract}
Numerical simulations of star cluster formation have advanced greatly during the past decade, covering increasingly massive gas clouds while accounting for more and more complex physics. In this review, I discuss the present state of the field, paying particular attention to the key physics that need to be included in cluster formation simulations. The main numerical techniques are summarized for a broad audience, before evaluating their application to the problem of cluster formation. A faithful reproduction of the observed characteristics of cluster formation can presently be achieved in numerical simulations. Ideally, this requires turbulent initial conditions to be combined with radiative feedback, protostellar outflows, and magnetic fields. With the exciting prospect in mind that our understanding of cluster formation will soon be revolutionized by  facilities like ALMA, JWST, and the EVLA, this review also identifies a number of areas that would particularly benefit from a joint observational and theoretical effort.
\end{abstract}

\section{\sc Introduction}
The clustered nature of star formation plays a central role in a wide range of astrophysical topics. Obtaining an understanding of the origin of the stellar initial mass function requires insight in the formation processes of coeval stellar populations, and hence an understanding of cluster formation \cite{klessen98,klessen00,bate03,bonnell03,bate09,smith08,bastian10,krumholz12,bressert12}. Stellar multiplicity is likely influenced by the clustered nature of star formation, as well as by the dynamical processing that occurs in young, dense clusters \cite{goodwin04,parker09,moeckel10,moeckel11b,bate12}. For similar reasons, the formation of planets and brown dwarfs does not occur in isolation, and the impact of their cluster environment has been the focus of much recent work \cite{adams04,whitworth06,olczak06,clarke07,bonnell08,bate12,parker12,dejuanovelar12,dukes12,thompson13}. On larger scales, clustered star formation may affect galactic-scale feedback and its influence on the interstellar medium (ISM) \cite{hopkins12,krause13}.

Of course, cluster formation is an interesting topic by itself too. Stellar clusters exist, and they have a wide range of ages and masses \cite{brodie06,portegieszwart10}. Some are almost as old as the Universe, and an understanding of cluster formation and its variation with the galactic environment may eventually lead us to understand how these ancient globular clusters formed \cite{harris94,elmegreen97,clarke00,kravtsov05,elmegreen10,kruijssen12c}.

In order to address the formation of clusters (and more specifically numerical simulations thereof), a definition of this process is in order. In this review, I will interpret `cluster formation' as the process of converting gas into stellar clusters. A key point here is that clusters are gravitationally bound by definition -- their unbound counterparts have been identified as associations for over half a century \cite{blaauw64}. The minimum number of stars in a cluster should be such that the system does not evaporate (or become unbound through other mechanisms) on a star formation time-scale. A conclusive estimate of this number requires more work \cite{moeckel12}, but a commonly adopted limit is $N>10^2$.

Cluster formation is governed by two simultaneous processes. Firstly, the gas needs to be converted into stars. Secondly, there needs to be some physical mechanism (or combination of mechanisms) at work that allows these stars to become gravitationally bound. Any self-contained numerical simulation of cluster formation needs to capture both of these processes. Note that the decomposition into these two main ingredients relies on the assumption that fragmentation occurs. The details of this requirement are not essential -- it is sufficient to assume that {\it some} fragmentation takes place, based on (for instance) a simple Jeans argument. In the remainder of this review, I therefore focus on the two aforementioned processes, and the exact distribution of mass over individual stars is not considered.

During the past decade, Moore's law has enabled a revolution in cluster formation simulations. The field has evolved from the initial simulations that included self-gravity, hydrodynamics, turbulence and a recipe for star formation \cite{klessen98,bate03,bonnell03}, to a new generation of simulations that account magnetic fields, radiative transfer, and a wealth of feedback mechanisms \cite{padoan11,dale12,krumholz12}. The increased degree of complexity obstructs a straightforward interpretation of these new results, because (as we shall see below) not all physics are independent.

This review focuses on the physical aspects of cluster formation simulations, and as such is aimed at observers and theorists alike. The aim is to identify the key physics, and establish a picture of how these are incorporated in current numerical simulations of cluster formation. The technical aspects of the simulations are not treated in much detail here, but there exist excellent reviews and code papers that the interested reader is referred to \cite{monaghan92,fryxell00,teyssier02,springel05c}.

\section{\sc The physics of cluster formation}
Figure~1 shows an HST image of 30~Doradus, a large and active star-forming region in the Large Magellanic Cloud. A single star-forming region (or complex) like this already reveals a wide range of environments in which star formation takes place. It is not hard to imagine that the various initial conditions affect the cluster formation process. In this section, I discuss the two mechanisms that were identified above as the key processes driving cluster formation, and I will focus on their invariance as well as their environmental dependence.

\subsection{\sc The conversion of gas into stars} \label{sec:sf}
As gas is consumed in the star formation process, the stellar mass increases. The advancement of this conversion is quantified as the star formation efficiency (SFE or $\epsilon$), which is given by
\begin{equation}
\label{eq:sfe}
\epsilon\equiv\frac{M_{\rm star}}{M_{\rm star}+M_{\rm gas}}=\frac{\epsilon_{\rm ff}}{t_{\rm ff}}t ,
\end{equation}
where $M_{\rm star}$ is the stellar mass, $M_{\rm gas}$ is the gas mass, $t$ is the time, and $\epsilon_{\rm ff}$ is the SFE per free-fall time $t_{\rm ff}$, i.e.~the mass fraction that is turned into stars every free-fall time, representing the rapidity of star formation. This leads to the crucial point that at a given time $t$, the SFE increases with the density as $\epsilon\propto t_{\rm ff}^{-1}\propto\rho^{1/2}$. If gravitational collapse were the only process governing star and cluster formation, then the SFE per-free fall time $\epsilon_{\rm ff}=1$ \cite{wang10}.
\begin{figure}
\label{fig:30dor}
\center\includegraphics[width=14cm]{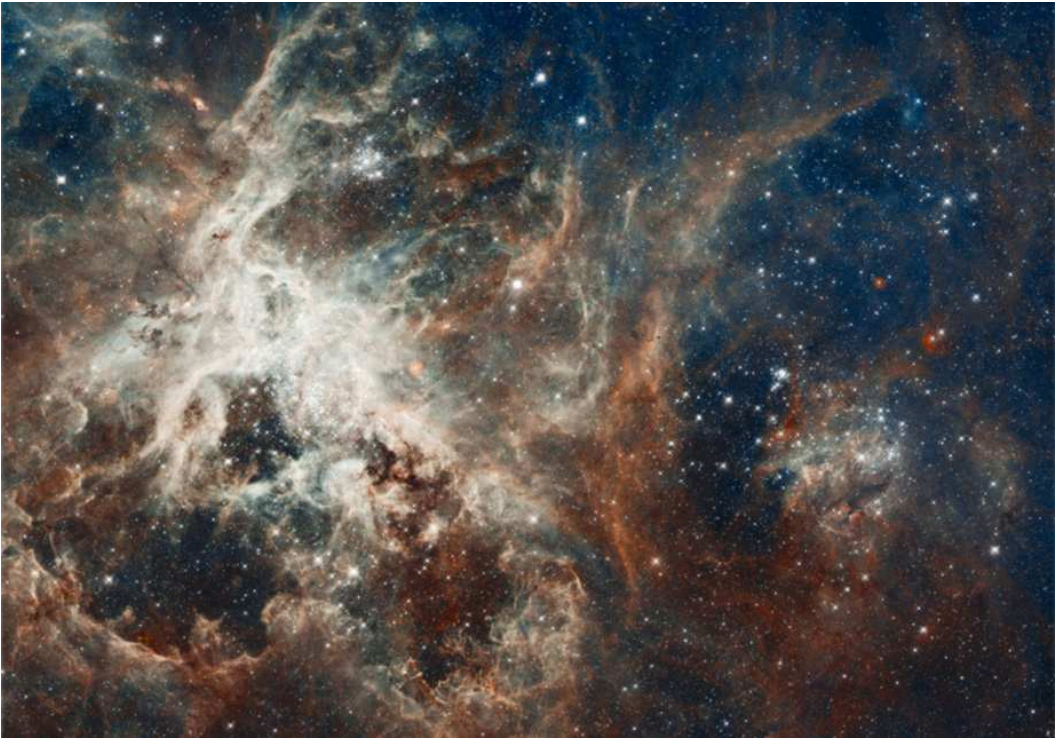}
\caption{Colour composite image of the star-forming region 30~Doradus. Red indicates H$\alpha$ emission, blue marks [O{\sc III}] emission, and green indicates a mix of both. The brightness is set by the HST/WFC3 F775W image. The dense, massive cluster R136 is located near the peak of the gas density, whereas more dispersed star formation takes place in lower-density regions. (Image credit: NASA/ESA/D.~Lennon et~al.)}
\end{figure}

The actual value of $\epsilon_{\rm ff}$ can be derived observationally, and shows a remarkably small variation with spatial scale \cite{krumholz07}. On galactic scales, the scaling relations for star formation yield $\epsilon_{\rm ff}=0.01$--$0.02$ \cite{kennicutt98b,elmegreen02,krumholz05}. This is remarkably inefficient compared to $\epsilon_{\rm ff}=1$ for pure gravitational collapse, which is possibly caused by feedback effects \cite{hopkins11}. On the pc-scales of star-forming regions in the solar neighbourhood, the result is very similar ($\epsilon_{\rm ff}=0.04$ \cite{evans09}), despite a change in spatial scale of four orders of magnitude. This is the regime relevant to cluster formation simulations.

The obvious (and well-known) conclusion is that the conversion of gas into stars and clusters is inefficient, and that gas may dominate the gravitational potential for several free-fall times. The eventual SFE is then set by the density (see equation~\ref{eq:sfe} above). With this empirical knowledge at hand, one can look at a giant molecular cloud (GMC) or star-forming region, and indicate the points where star formation will be most efficient. Returning to Figure~1, the young and dense, massive cluster R136 is visible towards the left-hand side of the image, where most of the dense gas is residing. More dispersed star formation is taking place in regions of lower gas density, which qualitatively underlines the importance of the density.

Numerical simulations of cluster formation are left with the challenge to bridge the gap between the SFE per free-fall time for pure gravitational collapse $\epsilon_{\rm ff}=1$ and the observed $\epsilon_{\rm ff}\sim0.02$ \cite{krumholz07}. While solving this issue is by itself already important for modelling the conversion from gas to stars, it is also essential in addressing the gravitational boundedness of young stellar structure (as we shall see in \S\ref{sec:bound}). There is a wide range of physical ingredients that may affect $\epsilon_{\rm ff}$, such as the initial density profile, turbulence, magnetic fields, feedback-related physics, and galactic-scale physics. The role of these aspects in cluster formation simulations is discussed in \S\ref{sec:sims}.

\subsection{\sc Gravitational boundedness} \label{sec:bound}
As indicated previously, a cluster formation simulation needs to include some physics to result in gravitational boundedness. In the classical picture of cluster formation, it depends on the SFE whether or not a system ends up being gravitationally bound \cite{tutukov78,hills80,lada84,adams00,geyer01,boily03,goodwin06,baumgardt07}. In this scenario, all stars form in clusters with a certain SFE $\epsilon$. A typical numerical simulation would initialize the stars being in virial equilibrium with a smooth background potential for the gas (which in some cases is  a scaled version of the stellar potential). The gas potential is then removed due to feedback (gradually or instantaneously), and the early evolution of the cluster is followed by modelling the dissipationless gravitational dynamics. Due to a SFE $\epsilon<1$, the cluster expands, possibly becoming unbound. The end result of this process is that only $5$--$10$\% of all stars ends up in gravitationally bound clusters \cite{lada03}. This numerical approach has been a very powerful tool in understanding what the requirements are for stellar structure to remain bound upon gas expulsion -- and hence built on from earlier analytical results \cite{hills80}. The key prediction of a gas-expulsion driven scenario for cluster formation, in which all stars initially form in clusters, is that young clusters should all go through a phase of accelerated expansion during which they respond to the loss of potential energy.

\begin{figure*} \label{fig:virialdis}
\center\includegraphics[width=8cm]{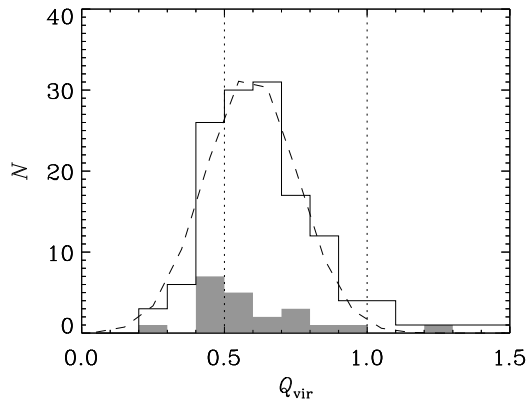}
\caption{The solid line shows a histogram of the virial ratios of the sub-clusters (ignoring the gravitational potential of the gas) from all snapshots of the Bonnell et al.~cluster formation simulation \cite{bonnell08}. The shaded histogram represents the set of sub-clusters from the last snapshot (after one free-fall time). The dashed line is a Gaussian fit to the data for all snapshots, with a mean value $Q_{\rm vir}=0.59$ and a standard deviation $\sigma_Q=0.16$. The vertical dotted lines indicate marginal gravitational boundedness ($Q_{\rm vir}=1$) and virial equilibrium ($Q_{\rm vir}=0.5$). Most sub-clusters are close to being virialized, indicating that they will not be strongly affected by gas expulsion. (Figure taken from \cite{kruijssen12}.)}
\end{figure*}
This classical picture of cluster formation has recently been challenged by new observational and numerical results. It has been realized that the Local Group young massive clusters (YMCs) NGC~3603 \cite{rochau10}, Westerlund~I \cite{cottaar12}, R136 \cite{henaultbrunet12} and the Arches \cite{clarkson12} are all in virial equilibrium, implying that they show no signs of violent gas removal and any resulting expansion. From the theoretical and numerical perspective, a similar prediction has been made at around the same time. Recall that at any given time $t$, the SFE increases with the density. This means that there has to be a density where the SFE becomes optimally efficient even before feedback starts playing a role, i.e.~most of the gas is consumed in the star formation process (modulo some mass loss by protostellar outflows). In an analysis of a smoothed particle hydrodynamics (SPH) simulation of cluster formation \cite{bonnell08}, we have indeed found gas-poor and virialized (sub-)clusters (see Figure~2 and \cite{kruijssen12}). In these systems, the gas accretion onto the sink particles can keep up with (or outruns) the global gas supply. This is not an artefact of the numerical method, because adaptive mesh refinement (AMR) simulations yield the same result \cite{girichidis12b} (see \S\ref{sec:num} below for a brief discussion on SPH and AMR methods).

The mechanisms leading to gas-poor (sub-)clusters proceed as follows.
\begin{enumerate}
\item[(1)]
Gas-poor cluster formation commences at the GMC stage, when no star formation has yet taken place. The gas velocity dispersion is a function of spatial scale and density, with the lowest dispersions occurring in small-scale, high-density cores \cite{larson81,maclow04}. As the first stars begin to form, they inherit the properties of the dense gas from which they condense, and hence obtain a lower velocity dispersion than is observed on the cloud scale \cite{offner09b}. In other words, the stars are initially sub-virial with respect to their parent GMC.
\item[(2)]
The stars decouple from the GMC and form a high-density, cluster core.
\item[(3)]
Due to the high density in the cluster core, the accretion of gas onto the sink particles and the accretion-induced shrinkage of the stellar system allow the local SFE to vastly exceed the global SFE of the parent GMC. Given sufficient density and time (in equation~\ref{eq:sfe} this requires $t\rho^{1/2}\propto t/t_{\rm ff}\gg1$), star formation thus leads to gas-poor and virialized clusters.
\end{enumerate}
These steps are three sides of the same coin, and are all essential for achieving the formation of bound clusters.

We are left with the interesting point that the entire cluster formation process is governed by the ISM density $\rho$ and the SFE per free-fall time $\epsilon_{\rm ff}$. A high density promotes bound cluster formation, while at a given density young stellar structure is more likely to remain bound if a high SFE is achieved on a short time-scale (i.e.~$\epsilon_{\rm ff}$ is large). If this is true, then the fraction of star formation occurring in bound stellar clusters (the cluster formation efficiency or CFE) should depend on the galactic environment. This has indeed been found in observational work \cite{goddard10,adamo11,silvavilla11}, and turns out to be quantitatively consistent with the above scenario for cluster formation \cite{kruijssen12d}. In Figure~3, I show the observed variation of the CFE (or $\Gamma$, \cite{bastian08}) with the galactic star formation rate density, together with the theoretically predicted variation if clusters form at the high-density end of the ISM density spectrum as detailed above. The figure shows that like star formation, cluster formation is inefficient. In a Milky Way-like galaxy, some 10\% of all star formation occurs in bound clusters, whereas at the highest densities this increases to $\Gamma=40$--$70$\%. The excellent agreement between the model and the observational data supports the identification of $\rho$ and $\epsilon_{\rm ff}$ as the main quantities driving cluster formation.

\subsection{\sc The role of numerical simulations} \label{sec:role}
\begin{figure*} \label{fig:cfe}
\center\includegraphics[width=8cm]{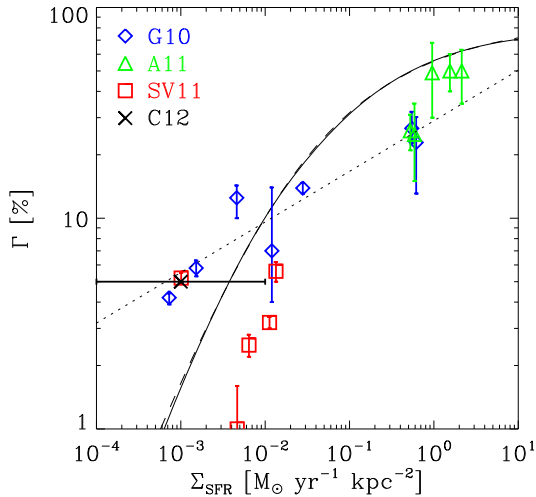}
\caption{CFE as a function of the star formation rate density. The symbols denote observed galaxies with $1\sigma$ error bars and indicate the samples from Goddard et al.~(\cite{goddard10}, blue diamonds), Adamo et al.~(\cite{adamo11}, green triangles) and Silva-Villa et al.~(\cite{silvavilla11}, red squares indicate their $\Gamma_{\rm MDD}$). The black cross indicates the integrated CFE of all dwarf galaxies from the sample of Cook et al.~\cite{cook12}, with a surface density range indicated by the horizontal error bar. The solid curve represents the `typical' theoretical  relation from \cite{kruijssen12d}, which is based on the idea that $\rho$ and $\epsilon_{\rm ff}$ drive cluster formation, and varies slightly for different galaxy and ISM properties. The dashed curve shows a simple analytical fit to the model (see equation~45 of \cite{kruijssen12d}), and the dotted line represents the original power law fit by Goddard et al.~\cite{goddard10}. (Figure taken from \cite{kruijssen12d}.)} 
\end{figure*}
It is clear that the density $\rho$ and the SFE per free-fall time $\epsilon_{\rm ff}$ play a central role in cluster formation, both in terms of the conversion of gas into stars and the evolution towards gravitational boundedness. The former of these two quantities is generally known -- observationally the density can be derived, and in numerical simulations it is set by the initial conditions. The SFE per free-fall time is the main unknown, and as explained in \S\ref{sec:sf}, theoretical and numerical studies of star formation are facing the challenge of resolving why observationally $\epsilon_{\rm ff}\sim0.02$ rather than the $\epsilon_{\rm ff}=1$ expected for pure gravitational collapse. This holds for cluster formation too -- if $\epsilon_{\rm ff}=1$, then all gas would be turned into stars in a single free-fall time. Because most star-forming regions evolve through more than one free-fall time \cite{krumholz07,evans09}, these stars would all end up being gravitationally bound. The key question is therefore not how to form a cluster, but how to {\it not} form one.

The problem at hand leaves us in the (un)fortunate situation that we know the necessary outcome of the simulations -- even quantitatively so. The formation of stars and clusters needs to be inefficient. Considering the wealth of physics that can be implemented in star formation simulations, there is a sincere risk (or luxury) that multiple mechanisms can be used to solve the problem. However, in the next section we shall see that cluster formation simulations are not (yet) affected by such a `first-world problem'.

\section{\sc Cluster formation simulations} \label{sec:sims}
Numerical simulations of cluster formation exist in a wide range of different flavours. In this section, I briefly summarize the advantages and limitations of the two main numerical methods, and discuss the influence of including various physical processes on the resulting efficiency of cluster formation.

\subsection{\sc Numerical methods: advantages and limitations} \label{sec:num}
There are two main numerical methods for simulating the evolution of astrophysical fluids in star formation. I briefly discuss them here -- the compilation of their advantages and disadvantages is by no means intended to be exhaustive, and the vibrant discussion in the literature has produced several more detailed discussions \cite{agertz07,wadsley08}.
\begin{enumerate}
\item[(1)]
{\it Smoothed Particle Hydrodynamics} (SPH) \cite{monaghan92,bate95,springel01,hubber11}. This method uses a particle representation of a fluid to solve the hydrodynamical equations of motion (the momentum, continuity, and thermal energy equations). Solving these equations requires the assumption of a certain equation of state that relates the density and pressure. SPH also requires certain assumptions regarding artificial viscosity and thermal conduction. The method employs particles to trace the gas mass, and whenever integrated quantities are calculated (such as the density), these properties are obtained by integrating them over the smoothing kernel -- a function that specifies the desired geometry of the spatial average. By allowing the radius of the kernel to vary with the density (to ensure the integration always occurs over a sufficient number of neighbouring particles), the spatial resolution becomes adaptive. The kernel can also be used to soften the gravitational forces (see below and the left-hand panel of Figure~4).

One of the main advantages of SPH is that is is Lagrangian: the particles carry the mass, and hence regions of enhanced density are automatically resolved at increased resolution. The obvious flip side is that in low-density regions, a large volume may be sampled by a single particle only. A second advantage of SPH is that it explicitly conserves angular momentum. This makes it a prime tool for studying rotationally supported or axisymmetric systems, such as protoplanetary discs or disc galaxies. Finally, the method is computationally relatively inexpensive, and the computation time scales very well for increasing numbers of processors -- provided that no star formation is taking place. A small number of particles on short time steps cannot efficiently be distributed over multiple processors.

Fundamentally, SPH always needs to overcome the problem that macroscopically, a real fluid is not particle-based. Solutions like the smoothing kernel and artificial viscosity are feasible compromises, but the relative freedom of their definition also introduces substantial uncertainty. Perhaps most importantly, SPH suffers from its inability to resolve fluid discontinuities, like phase separations or mixing. This is a direct result of its particle-based nature, and while these and other problems can be offset to some degree by increasing the resolution, adapting the smoothing kernel \cite{dehnen12}, changing the artificial viscosity \cite{read12}, or modifying the equations of motion \cite{hopkins13}, they cannot be alleviated fully. Another promising avenue is Godunov SPH, which solves the Riemann problem for all particle neighbour pairs \cite{inutsuka02}.
\begin{figure*} \label{fig:num}
\center\includegraphics[width=8.7cm]{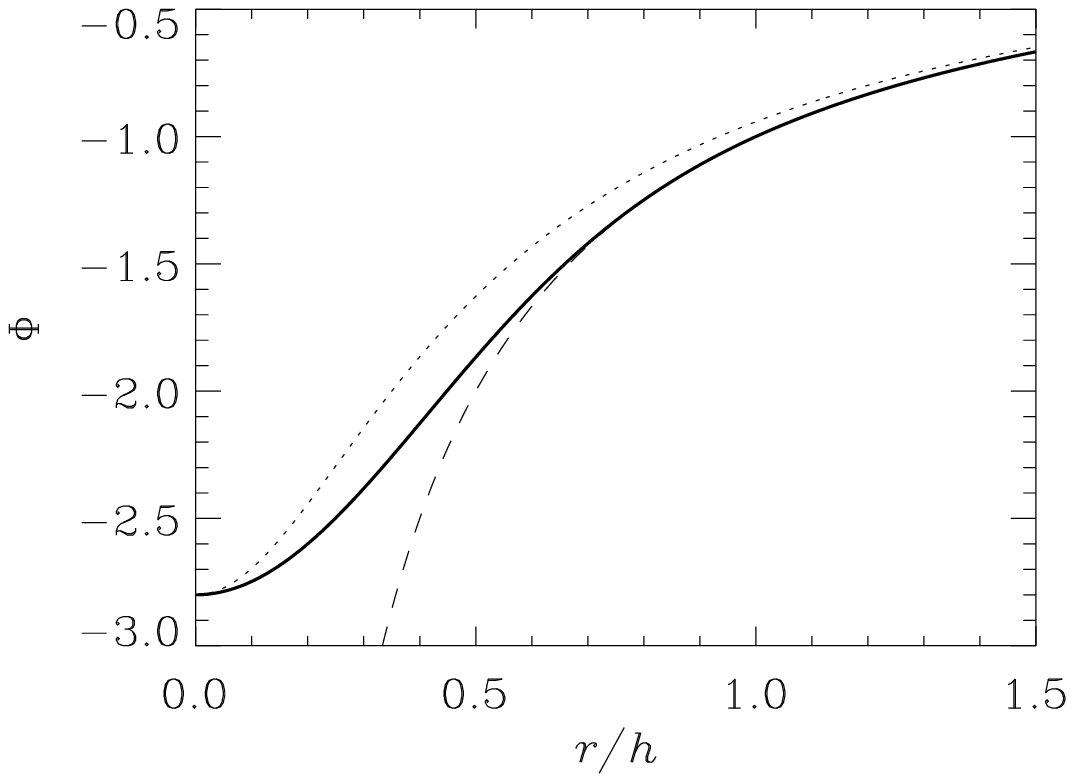}\includegraphics[width=7cm]{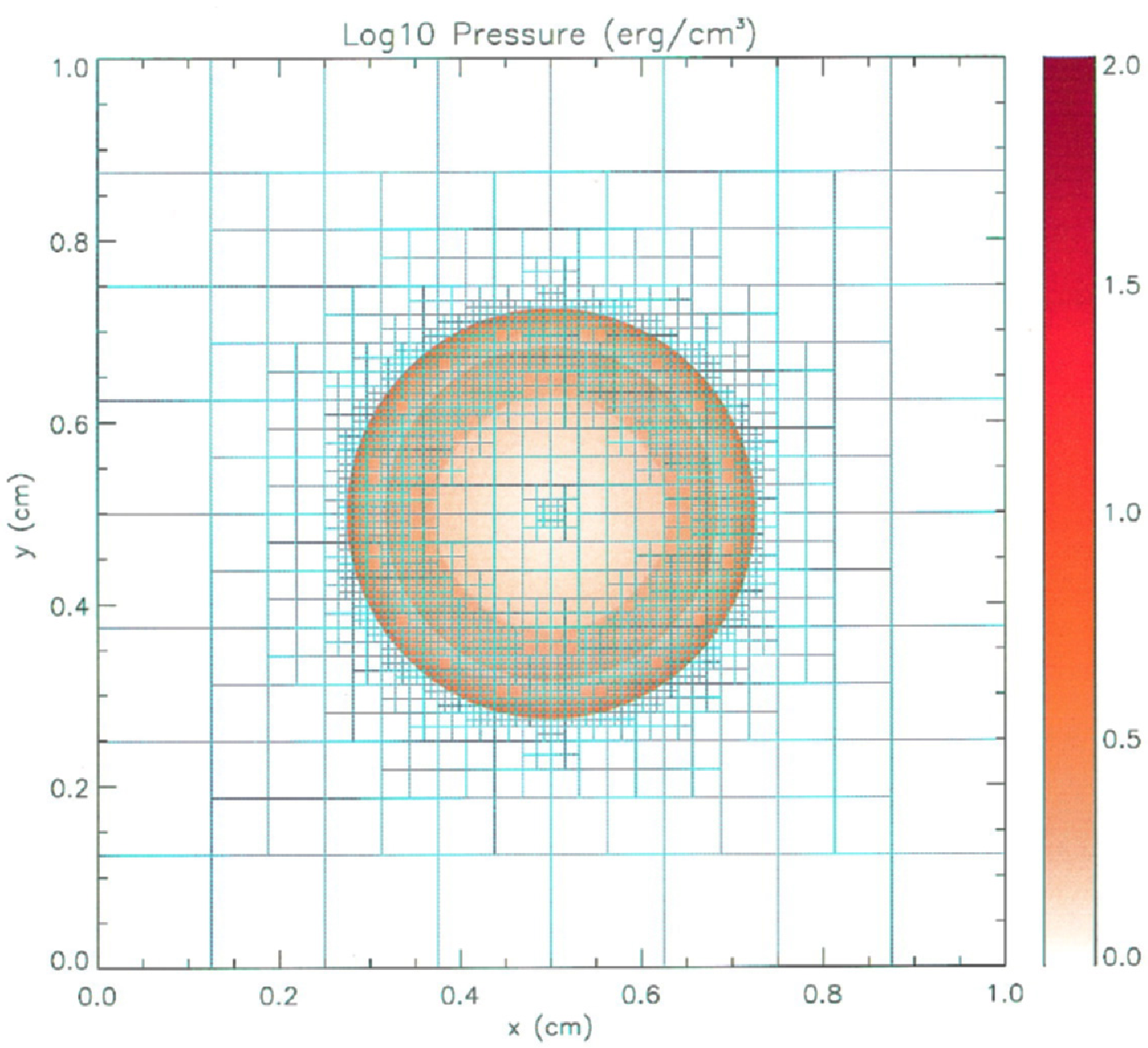}
\caption{{\it Left}: Gravitational potential as a function of normalized radius. The dashed line shows a Newtonian point-mass potential, whereas the solid and dashed line illustrate the effect of a softened potential using a spline kernel (solid) or a Plummer profile (dotted, with scale radius $r_{\rm p}=h/2.8$). The force divergence at small radii is removed by using softened gravity. (After Figure~13 of \cite{springel01}.) {\it Right}: Adaptive Mesh Refinement grid produced by the {\sc Flash} code \cite{fryxell00} for a Sedov blast wave test. The mesh uses eight levels of refinement, and the colour map indicates the gas pressure. Note the substantial resolution increase near regions of high density, as well as near the centre of the explosion. (Figure taken from \cite{fryxell00}, reproduced by permission of the AAS.)}
\end{figure*}

\item[(2)]
{\it Adaptive Mesh Refinement} (AMR) \cite{kravtsov97,fryxell00,teyssier02,oshea04}. This method employs a grid-based approach to solve the hydrodynamical equations of motion. Traditional grid codes use a static grid of cells, and the relevant physical quantities to describe the state of the gas are defined for each cell. Because the cells themselves are static, the motion of gas is achieved by shifting gas mass (along with its properties) from cell to cell. The AMR technique vastly improves this approach by dividing up the cells to add spatial resolution where necessary (e.g.~in regions of high mass density). An example is given in the right-hand panel of Figure~4, where the adaptive refinement is illustrated with a Sedov blast wave test. As in SPH, the pressure is obtained from the energy and density by using an equation of state, and AMR also employs artificial viscosity to capture shocks. Note that in order to include stars (or sinks), AMR methods still have to include particles like SPH codes do. The gravitational interaction between grid cells and particles is generally solved using a particle-mesh technique \cite{hockney81,efstathiou85}.

The AMR method is Eulerian: the mass is carried by volume elements. An important implication of this approach is that AMR codes are exquisite tools for modelling fluid discontinuities, which are well-resolved thanks to the intrinsically discontinuous nature of the mesh. A cell-based method is naturally appropriate to describe a fluid, even though interpolation between different cells is necessary to accurately describe gradients without increasing the spatial resolution.

The fact that AMR is Eulerian is also a disadvantage. Angular momentum is not exactly conserved on a cartesian grid, and the method is therefore less suitable for modelling rotationally supported systems. The grid also introduces a special orientation in simulated systems, which can lead to the spurious alignment of structures. 
\end{enumerate}

In both methods, star formation occurs by spawning `sink' particles \cite{bate95} at positions where the gas satisfies a certain set of requirements. The specific conditions vary between different codes, but some examples are \cite{federrath10,hubber13}:
\begin{enumerate}
\item[(1)]
The gas density exceeds some critical density threshold.
\item[(2)]
The newly spawned sink particle does not overlap with any other sink particle.
\item[(3)]
The gravitational potential has a local minimum.
\item[(4)]
The divergence of the gas velocity field is smaller than zero (i.e.~a converging flow).
\item[(5)]
The gas is Jeans-unstable and/or gravitationally bound.
\end{enumerate}
Of these examples, the top three form a set of conditions that gives reliable results \cite{hubber13}. After having spawned, sink particles can continue to grow in mass by accreting gas \cite{bate95}. The conditions for gas accretion vary. In some codes, it is required that the gas resides within a certain radius $r_1$ and is gravitationally bound to the sink particle, whereas within a radius $r_2<r_1$ the gas is accreted regardless of its dynamical state \cite{bonnell08}. Alternatively, a physically motivated time-scale for gas situated within the sink particle radius can also be adopted \cite{hubber13}. Using a friends-of-friends algorithm, it is also possible to allow sink particles to merge \cite{krumholz04}. As is done for the gas, the gravity between the sink particles is generally softened (see Figure~4). This is done to prevent spurious two-body effects with the gas, and to account for the intrinsically extended nature of the sinks. However, the obvious downside for collisional systems like stellar clusters is that the long-term $N$-body dynamics are not accurately modelled.

The advantage of having more than a single numerical method at hand is that it is possible to verify whether results are numerical artefacts or are real. The particle-based and grid-based methods have recently also been combined in the moving-mesh code {\sc arepo} \cite{springel10}, and the first applications to cluster formation have already appeared in the literature \cite{greif11}. While the concept of a moving mesh introduces its own technical challenges, a broader application of the method to star and cluster formation problems should be expected in the near future.

The cluster mass and time-scales that can currently be modelled depend on the numerical method and the physics that are included. All other things being equal, an AMR code that includes magnetic fields will be able to follow the formation of a less massive cluster than an SPH code that only includes the essential hydrodynamics and gravity solvers. Having said that, present state-of-the art facilities and codes are capable of reliably tracking the formation of a cluster from a $M_{\rm gas}\leq10^4~\msun$ cloud for $\sim1$ free-fall time, with a resolution that is sufficient to resolve the hydrogen burning limit at $m=0.08~\msun$ \cite{tilley07,bonnell08,krumholz12}. If the low-mass end of the stellar mass function is not relevant to the problem at hand, then a larger mass range can be covered \cite{dale12}.

\subsection{\sc Input physics} \label{sec:physics}
The conversion of a real gas cloud into a stellar cluster is governed by a wide range of physical mechanisms, and in order to provide a simplified model of this process, numerical simulations have to contain descriptions for some subset of these mechanisms. As indicated above, the SFE per free-fall time $\epsilon_{\rm ff}$ is of order unity in a hypothetical cloud that experiences no physics other than gravity and basic hydrodynamics. If this were true in nature, all stars would form in gravitationally bound clusters. The main goal of this section is to identify which physics decrease $\epsilon_{\rm ff}=1$ to the necessary $\epsilon_{\rm ff}\sim0.02$, and hence make cluster formation inefficient. Next to addressing that question, it also serves as a showcase of what  can currently be achieved with state-of-the-art cluster formation simulations.

\subsubsection{\sc The initial density profile and turbulence} \label{sec:initial}
Using both using SPH and AMR codes, it has been shown that the initial cloud density profile plays a relatively minor role in setting $\epsilon_{\rm ff}$, and only affects the degree of substructure during gravitational collapse \cite{bate09b,girichidis11}. This is illustrated in Figure~5, which shows a column density map of three small-$N$ cluster formation simulations by \cite{girichidis11}. The simulations each track the evolution of a collapsing gas cloud, which initially follows a top-hat profile, a Bonnor-Ebert sphere \cite{ebert55,bonnor56}, or a self-similar power law profile. After $0.6$--$0.9$ free-fall times, the simulations have reached a comparable SFE, and only differ in terms of their substructure. Because these systems are not initially supervirial, the differences in substructure will eventually be erased, leaving three stellar clusters with comparable properties.
\begin{figure*} \label{fig:girichidis}
\center\includegraphics[width=\textwidth]{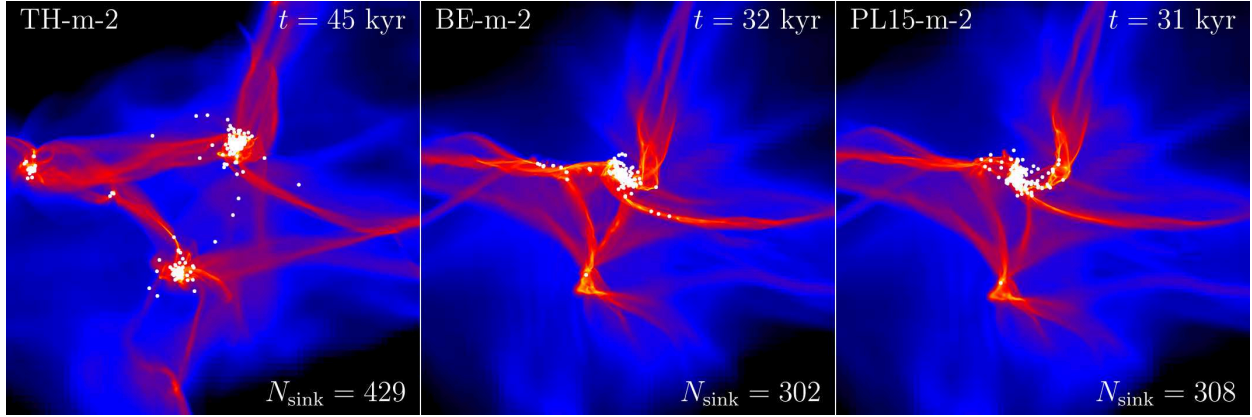}
\caption{From left to right, the panels show column density maps of three cluster formation simulations by \cite{girichidis11}, initialized ass a top-hat density profile, a Bonnor-Ebert sphere, and a self-similar power law profile, respectively. The white dots indicate sink particles, and the number of sinks is indicated in the bottom-right corner of each panel. (Figure adapted from \cite{girichidis11}.)}
\end{figure*}

The internal energy of cold ($T=10$--$20$~K) GMCs is dominated by supersonic turbulence \cite{maclow04,mckee07}. The time-scale for the dissipation of the turbulent energy is typically a crossing time, i.e.~1--2 free-fall times. As a result, an initial turbulence spectrum should slow down the collapse towards star formation. In cluster formation simulations, turbulence can either be initialized and left to decay, or be maintained through turbulence forcing \cite{federrath10}. Most current estimates of the effect of turbulence on $\epsilon_{\rm ff}$ do not employ forcing, which is physically reasonable, because a gravitationally collapsing gas cloud is probably dynamically decoupled from whatever is driving the turbulence on a larger scale. Some of the largest cluster formation simulations to date (up to $10^4~\msun$, using SPH and AMR), allow a straightforward estimate of $\epsilon_{\rm ff}$. In these simulations, the SFE per free-fall time decreases from $\epsilon_{\rm ff}=1$ to $\epsilon_{\rm ff}\sim0.35$ due to the inclusion of an initial turbulence spectrum \cite{bate03,bonnell08,krumholz12}. Note that the variation of $\epsilon_{\rm ff}$ due to a different shape of the initial turbulence spectrum (i.e.~different $n$ in $P(k)\propto k^{-n}$, where $k$ is the wavenumber) is negligible \cite{bate09b}.

\subsubsection{\sc Protostellar outflows} \label{sec:outflows}
\begin{figure*} \label{fig:hansendale}
\center\includegraphics[width=8.9cm]{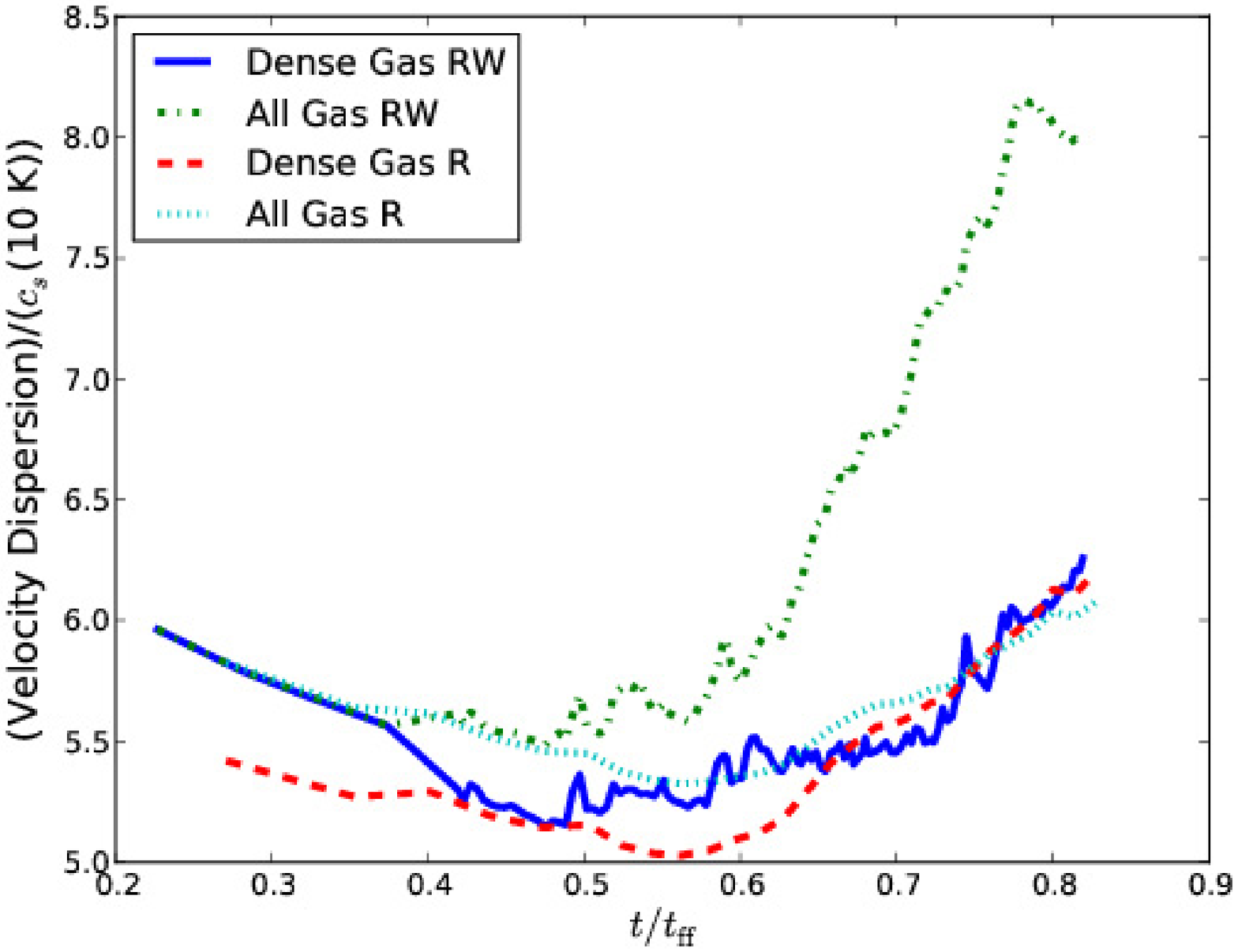}\includegraphics[width=7.1cm]{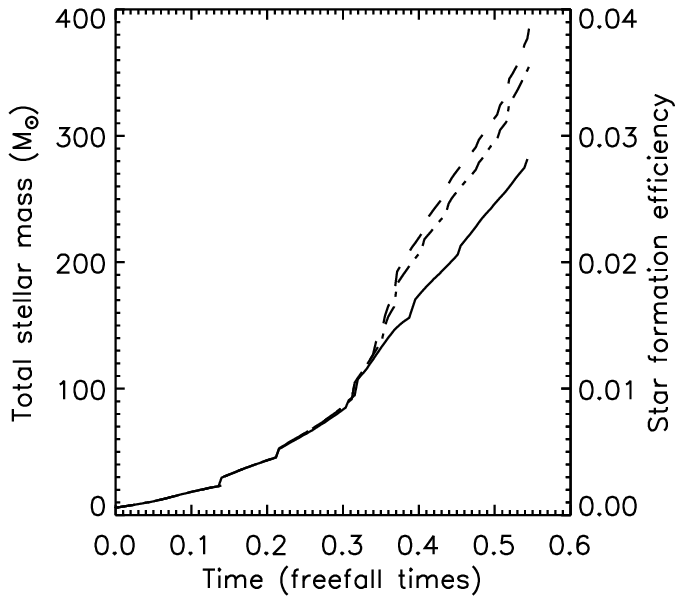}
\caption{{\it Left}: Effect of protostellar outflows on the Mach number of the gas. The figure shows the Mach number of different (parts of) simulations by \cite{hansen12} as a function of time. The label `RW' in the legend indicates that radiative transfer and outflows are both included, whereas `R' indicates that only radiative transfer is accounted for. The similarity of the curves representing the dense gas implies that the energy injected by the outflows only affects low-density gas. (Figure taken from \cite{hansen12}, reproduced by permission of the AAS.) {\it Right}: Stellar mass (or SFE) evolution of a simulation by \cite{dale07b} that includes photoionizing radiation from an external radiation source. The solid line indicates a model without feedback, whereas the dashed line includes photoionization. The dash-dotted line excludes protostellar cores that have been directly triggered by the photoionizing radiation. (Figure taken from \cite{dale07}.)}
\end{figure*}
The previous section considered the cascade of turbulent energy down to smaller scales, but in a star-forming environment, energy is also generated on small scales by (proto)stellar feedback. Both the effects of protostellar outflows and stellar winds have been treated independently in cluster formation simulations. \cite{hansen12} use AMR simulations that include protostellar outflows (with and without radiative transfer) to address their influence on $\epsilon_{\rm ff}$. They find that irrespective of radiation physics, the energy injected by protostellar outflows does not strongly affect the rapidity of star formation, despite the outflows being able to replenish the turbulent energy (also see \cite{nakamura07}). The outflows do unbind a substantial fraction of the mass and hence limit the maximum SFE to $\epsilon_{\rm max}\sim1/3$, but as discussed by \cite{krumholz12} this result strongly depends on the initial conditions (e.g.~the binding energy of the cloud), which fundamentally complicates a population-averaged estimate of how important outflows really are in promoting inefficient cluster formation.

The reason that the $\epsilon_{\rm ff}$ is weakly affected by outflows in the simulations of \cite{hansen12}, is that most of the mass and energy loss occurs through the path of least resistance. In the absence of additional physics (see \S\ref{sec:coll} below), this leaves the global cloud dynamics unchanged, nor does it influence the dense gas. The left-hand panel of Figure~6 shows this quantitatively for the simulations including radiative transfer. The turbulent energy of {\it all} gas increases by almost a factor of two due to outflows ($E_{\rm turb}\propto{\cal M}^2$, with ${\cal M}$ the Mach number), whereas for the dense gas it is unaffected by the energy injection.

\subsubsection{\sc Photoionizing radiation} \label{sec:ion}
Photoionization injects thermal energy into the ISM surrounding the ionization source. Provided that the energy coupling to the star-forming gas is efficient, the resulting H{\sc ii} regions can affect the rapidity of cluster formation. The influence of photoionizing radiation on star formation has been studied using SPH \cite{dale05,dale07b,dale11} and AMR methods \cite{peters10}. The main outcome of these simulations is that the dense gas is often weakly affected by ionizing radiation due to self-shielding, and hence the feedback energy escapes along the path of least resistance (as is the case for protostellar outflows, see \S\ref{sec:outflows}), filling the voids and pseudobubbles generated by the turbulent velocity field \cite{dale11}. 

The SFE per free-fall time is almost unaltered by photoionization. This is illustrated for the simulation by \cite{dale07} in the right-hand panel of Figure~6, where we see that an external source of photoionizing radiation in fact slightly increases the SFE per free-fall time, from $\epsilon_{\rm ff}\sim0.08$ to $\epsilon_{\rm ff}\sim0.1$. The increase is caused by triggered star formation, but this effect largely disappears in later simulations that include photoionization as a self-consistent, internal feedback mechanism that originates from multiple sources in the simulation \cite{dale11}. In low-mass clouds, the thermal energy injected by photoionization is capable of exceeding the binding energy of the cloud (i.e.~the ionized sound speed exceeds the cloud escape velocity \cite{bressert12b}), and hence a substantial fraction of the gas may be unbound \cite{dale13}. The resulting porosity of the ambient ISM may render it less susceptible to feedback processes acting on a longer time-scale, such as supernovae (see \S\ref{sec:winds}).

\subsubsection{\sc Radiative transfer and feedback} \label{sec:rad}
While photoionizing feedback strictly refers to the thermal effects of the ionization of gas by EUV or X-ray photons, radiative feedback in general concerns the transfer of energy from the radiation field to the gas via photodissociation, dust heating or radiation pressure, and hence includes thermal and kinetic (pressure) effects. It is conceivable that contrary to photoionization, a broader range of radiative feedback does affect the efficiency of star and cluster formation. Simple analytical arguments already show that radiation pressure is the dominant source of feedback in clusters with masses $M\geq10^4~\msun$ if the surface density exceeds $\Sigma\sim500~\msun~{\rm pc}^{-2}$, which implies a radius of
\begin{equation}
\label{eq:rrad}
R\leq2.5~{\rm pc}\times(M/10^4~\msun)^{1/2}~{\rm with}~M\geq10^4~\msun,
\end{equation}
for radiative feedback to be important \cite{fall10}. Note that this does not mean that radiative feedback is negligible at lower masses, and the interplay between different feedback mechanisms likely determines to what extent it affects $\epsilon_{\rm ff}$ (see \S\ref{sec:coll}).

SPH and AMR simulations agree in that radiative feedback inhibits fragmentation and the subsequent formation of low-mass stars \cite{offner09,bate09,krumholz12}, because the temperature increase boosts the Jeans mass as $M_{\rm J}\propto T^{3/2}\rho^{-1/2}$. The effect is quantified in Figure~7, which shows the time-evolution of two small-$N$ AMR simulations with and without radiative feedback \cite{offner09}, as well as stellar mass functions resulting from both runs. Integration of these shows that accounting for radiative feedback decreases the SFE per free-fall time by about a factor of two, while the time-sequence in the left-hand panel shows that the run including radiative feedback retains a higher degree of substructure. Of course, these effects should only occur if the cluster is sufficiently massive to host massive stars. For the mass and radius constraints of equation~\ref{eq:rrad}, it is clear that this condition is satisfied in the regime where radiative feedback dominates over other feedback mechanisms.
\begin{figure*} \label{fig:offner}
\center\includegraphics[width=9.1cm]{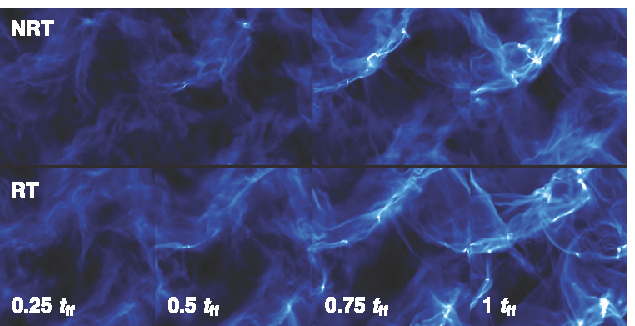}\includegraphics[width=6.9cm]{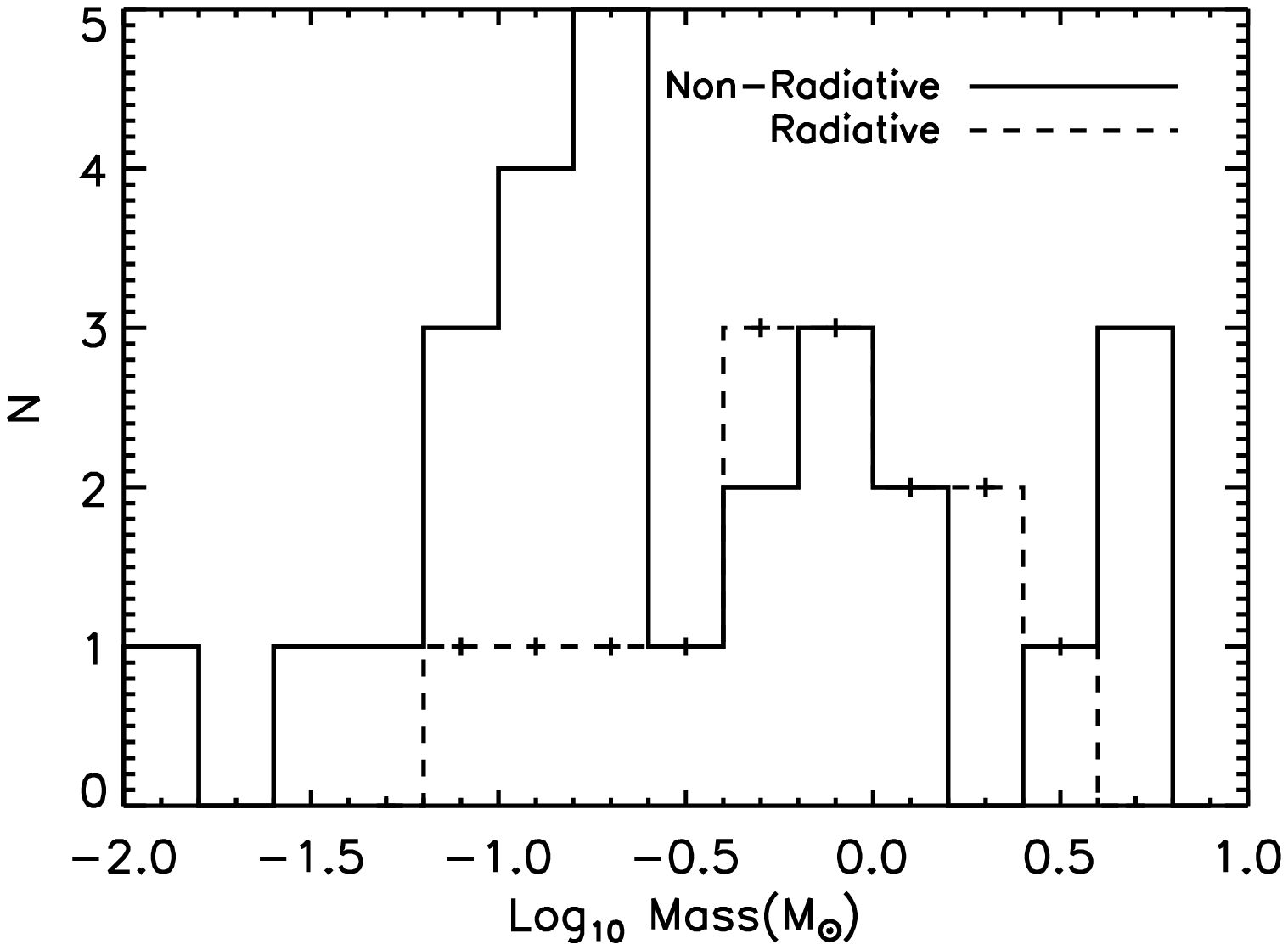}
\caption{{\it Left}: Gas column density evolution of two simulations by \cite{offner09}. The top row (NRT) does not include radiative feedback, whereas the bottom row (RT) does. (Figure adapted from \cite{offner09}, reproduced by permission of the AAS.) {\it Right}: Mass function including the disc and star particle masses for the two simulations in the left-hand panel. (Figure taken from \cite{offner09}, reproduced by permission of the AAS.)}
\end{figure*}

\subsubsection{\sc Stellar winds and supernovae} \label{sec:winds}
It is not straightforward to quantify the influence of stellar winds or supernovae on the efficiency of cluster formation, because their impact may be delayed by some absolute time-scale. As a result, their effectiveness depends on the density of the star-forming region, or more specifically on the number of free-fall times a region undergoes before wind and supernova feedback become important. In the high-density star-forming regions that presumably lead to the formation of massive, dense clusters, the energy dissipation time-scale is much shorter than the free-fall time, and hence the gas may be used up in star formation before the onset of wind and  supernova feedback, or the injected energy is dissipated \cite{murray10}. However, in regions of lower density, the details of the sequence of wind and supernova feedback determine whether either mechanism is capable of affecting the efficiency of star and cluster formation \cite{pelupessy12}.

Once stellar wind feedback starts, it can have a substantial effect on $\epsilon_{\rm ff}$ if gas is still present. The SPH simulation of \cite{dale08} shows that depending on the details of the wind (e.g.~isotropic or collimated), it can decrease the SFE per free-fall time by up to a factor of two, from $\epsilon_{\rm ff}\sim0.2$ to $\epsilon_{\rm ff}\sim0.1$. The influence of collimated winds is delayed with respect to isotropic winds, but eventually their large-scale impact starts to influence the accretion process onto the feedback sources, and both wind types become equally effective in inhibiting cluster formation.

While supernova explosions are the main feedback mechanism on galactic scales \cite{mckee77}, the relatively long delay time for supernovae implies that supernova feedback typically occurs after other feedback mechanisms have already done their damage to the local ambient medium. If these feedback processes have been able to remove the gas, there is no star formation left to be slowed down. It was shown by \cite{pelupessy12} how the sequence and relative strength of the different feedback mechanisms is crucial in setting the importance of supernovae. If the impact of preceding feedback was weak, then supernovae simply halt star formation altogether \cite{pelupessy12}, provided that the region is of sufficiently low density to expel the gas before the energy is dissipated \cite{murray10}.

\begin{figure*} \label{fig:pelupessy}
\center\includegraphics[width=8cm]{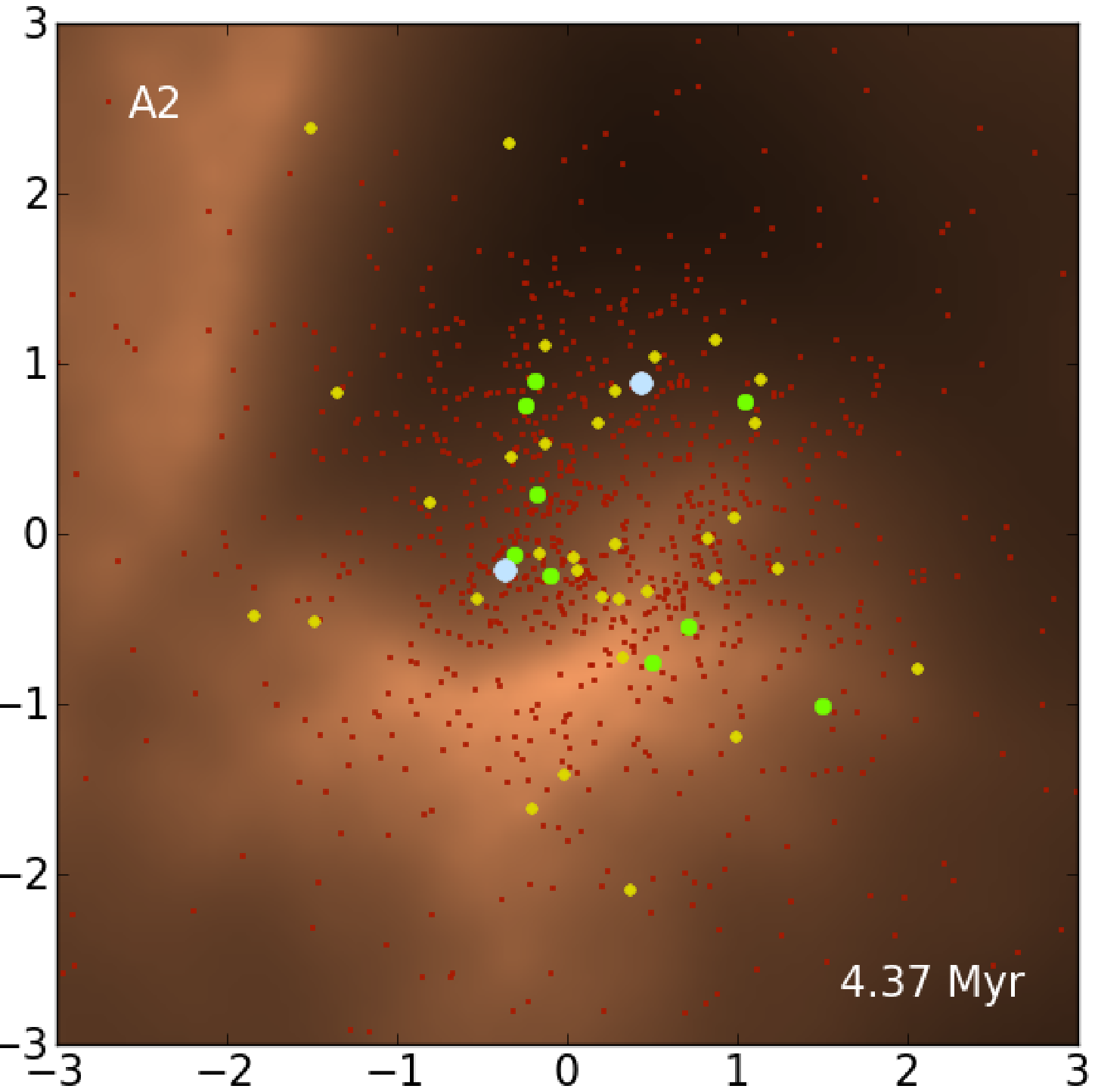}\includegraphics[width=8cm]{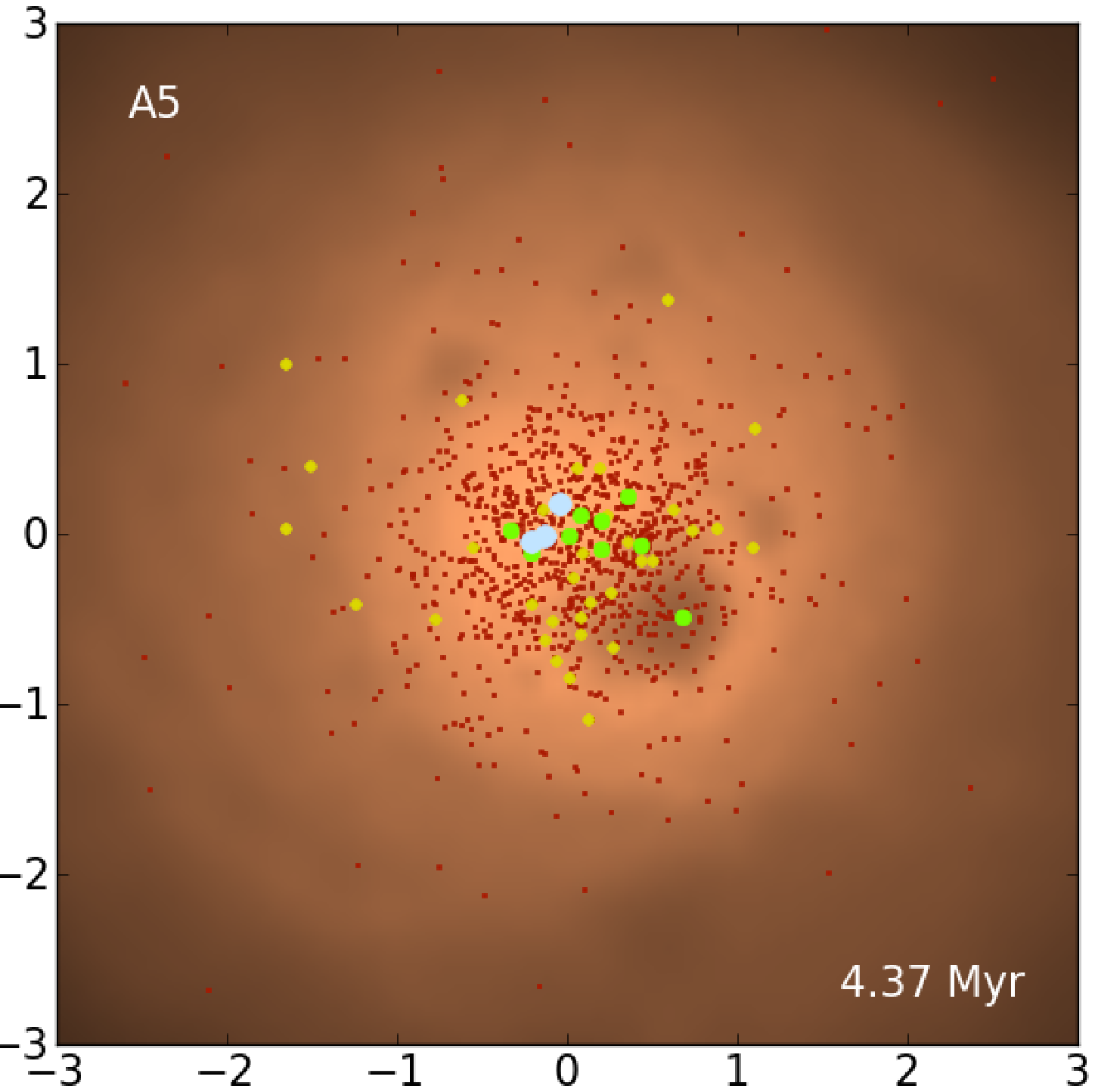}
\caption{Distribution of gas and stars (dots) in two simulations of stellar wind and supernova feedback by \cite{pelupessy12}. The left-hand panel shows the system when the feedback efficiency (see text) is $f_{\rm fb}=0.1$, whereas in the right-hand panel $f_{\rm fb}=0.01$. The strong stellar winds in the high-$f_{\rm fb}$ run unbind the gas before the first supernova (which occurs at $t\sim10$~Myr), whereas in the low-$f_{\rm fb}$ run about half the gas mass is unbound by the first supernova. (Figure adapted from \cite{pelupessy12}.)}
\end{figure*}
The sensitivity to the feedback history is illustrated in Figure~8, which shows two SPH simulations of a small-$N$ cluster embedded in its parent gas cloud. While the simulations do not include the formation of sink particles, the self-consistent hydrodynamical modelling of the gas and stars allow them to evolve independently, which is a crucial step compared to modelling the pure $N$-body dynamics. The simulations have feedback efficiencies $f_{\rm fb}$ (a free parameter indicating the efficiency of energy coupling between the feedback and the ISM) that differ by an order of magnitude. The high-$f_{\rm fb}$ run develops bubbles and outflows, whereas the low-$f_{\rm fb}$ run remains relatively quiescent and more gas-rich. By the time the first supernova explodes, there is no gas left in the high-$f_{\rm fb}$ simulation, whereas in the low-$f_{\rm fb}$ run the gas and stellar masses are comparable and all remaining gas is unbound by the supernova. While this does not cause a major difference in the eventual boundedness of the clusters (the bound fraction is $f_{\rm bound}\sim 0.6$ in both cases), other simulations with higher initial gas fractions (i.e.~lower SFEs) do yield unbound clusters after the first supernova, caused by the sudden removal of the gravitational potential.

In conclusion, the influence of feedback by stellar winds and supernovae once again reduces to a question of density -- how does the ambient ISM evolve before either feedback mechanism becomes important? In high-density environments, the evolution of a region prior to stellar evolutionary feedback can be substantial and hence it should not play an important role (even when ignoring the enhanced energy dissipation rate at high density). The primary consequence is that at high densities very high SFEs can be reached. At lower densities, the impact can be dramatic, provided that the SFE is so low that the region is still producing stars after several Myr. In such a case, supernovae and winds can halt star formation altogether, and mark the key moment in time at which the long-term gravitational boundedness of stellar structure is determined.

\subsubsection{\sc Magnetic fields} \label{sec:magn}
Magnetic fields can provide support to low-density gas on the spatial scales of clouds, thus inhibiting their collapse \cite{price09,wang10,padoan11,padoan12}. Such a large-scale influence of magnetic fields largely complements the effect of radiative feedback, which primarily leads to heating on smaller scales \cite{price09}. Numerical simulations including magnetic fields have been performed using SPH and AMR methods, and despite the numerical differences, the main results are consistent.

\begin{figure*} \label{fig:padoan}
\center\includegraphics[width=6.9cm]{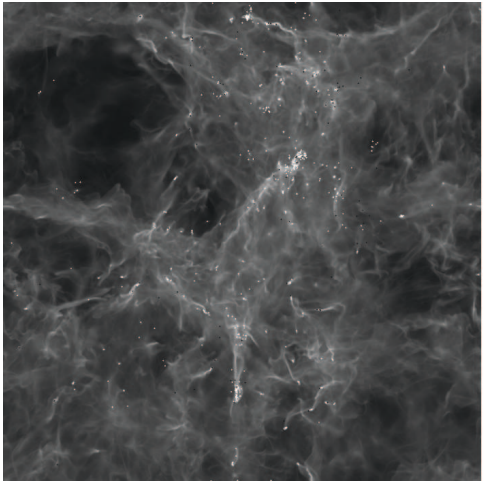}\includegraphics[width=9.1cm]{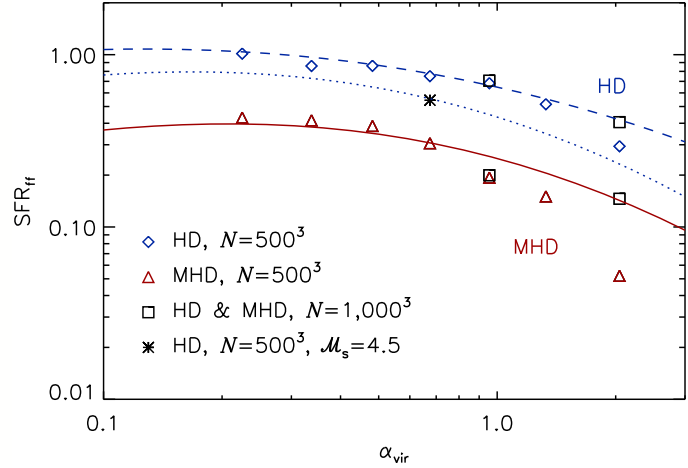}
\caption{{\it Left}: Column density map of a magnetohydrodynamic star formation simulation by \cite{padoan11}, at a time where a SFE of $\epsilon\sim0.1$ is reached. White dots indicate stellar-mass sink particles, whereas black dots represent sub-stellar sink particles. {\it Right}: SFE per free-fall time $\epsilon_{\rm ff}$ as a function of the virial parameter $\alpha_{\rm vir}$ for different Mach numbers and magnetic field strengths. Symbols indicate various simulations like the one shown in the left-hand panel, but with different initial conditions, and the lines indicate the results of an analytic model. Models including magnetic fields are shown in red (MHD), whereas purely hydrodynamical results are shown in blue (HD). The Mach number corresponding to the dashed line is twice as high as that of the dotted line, and matches that of the MHD results. For the adopted sound speed and Alfv\'{e}nic velocity, the presence of magnetic fields suppresses the SFE per free-fall time by a factor of $\epsilon_{\rm ff,HD}/\epsilon_{\rm ff,MHD}\sim3$. (Both panels taken from \cite{padoan11}, reproduced by permission of the AAS.)}
\end{figure*}
Depending on the details of the initial conditions (such as the magnetic field strength and the cloud virial parameter), the presence of magnetic fields can slow down star formation by a factor of a few. For $\beta\equiv2c_{\rm s}^2/v_{\rm A}^2=0.39$, with $c_{\rm s}$ the sound speed and $v_{\rm A}$ the Alfv\'{e}nic velocity, which depends on the magnetic field strength $B$ as $v_{\rm A}\propto B\rho^{-1/2}$, \cite{padoan11} find that $\epsilon_{\rm ff}$ is reduced by a factor of three. The precise value of this decrease depends on the Mach number and the virial parameter, as illustrated in Figure~9. Like for the inclusion of magnetic fields, the variation of $\epsilon_{\rm ff}$ for a plausible range of virial parameters ($\alpha_{\rm vir}=0.3$--$3$) also covers a factor of three, and shows that a slightly supervirial nature of clouds could inhibit star and cluster formation to a similar degree as magnetic fields. However, GMCs are observed to be roughly in virial equilibrium \cite{bolatto08,heyer09}, and hence the impact of $\alpha_{\rm vir}$ cannot be too substantial.

\subsubsection{\sc External effects} \label{sec:ext}
All of the cluster formation simulations discussed so far considered a gas cloud in isolation. However, some of the most famous cluster formation sites are nowhere near quiescent, and are induced by large-scale gas inflows, galaxy mergers, or other galactic events \cite{holtzman92,whitmore99,adamo11}. Could it be that cluster formation proceeds differently in converging or compressed gas flows? Is it accelerated or slowed down? At present, it is  reasonable to neglect galaxy-scale effects, simply because a faithful numerical simulation of star cluster formation is a major challenge by itself. However, it should also be kept in mind that galactic processes could play a role.

\begin{figure*} \label{fig:maps}
\includegraphics[width=\textwidth]{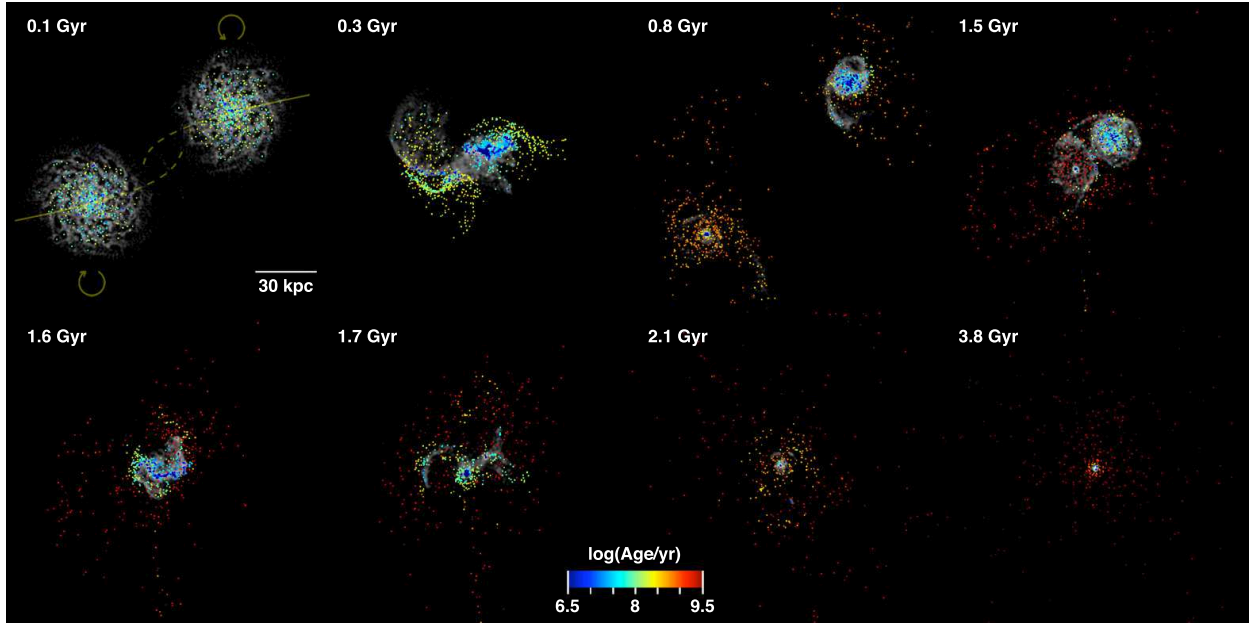}
\caption{Evolution of the star cluster population during a galaxy merger simulation from \cite{kruijssen11}. The surface density of the gas is displayed in greyscale, while the particles that contain star clusters are shown in colours denoting the ages of the clusters as indicated by the legend. The subsequent images show the collision at eight characteristic moments: $t=0.1$~Gyr, briefly before the first passage; $t=0.3$~Gyr, just after the first passage; $t=0.8$~Gyr, in between the first and second passage; $t=1.5$~Gyr, just before the second passage; $t=1.6$~Gyr, just after the second passage; $t=1.7$~Gyr just before the final coalescence; $t=2.1$~Gyr, during the coalescence and the infall of remaining gas clouds; $t=3.8$~Gyr, when only a merger remnant is left. There is a wide range of evidence suggesting that cluster formation and destruction proceed differently in the high-density ISM than in a quiescent galaxy disc. (Figure taken from \cite{kruijssen12c}.)}
\end{figure*}
The dependence of the cluster formation process on the density implies that the CFE (the fraction of all star formation occurring in gravitationally bound clusters) varies with the galactic environment (see \S\ref{sec:bound} and Figure~3), simply because the properties of the parent GMCs do \cite{kruijssen12d}. But less obvious galaxy-scale physics are also capable of influencing cluster formation. In dwarf galaxies, the short dynamical friction time-scales can drive the assembly of a nuclear cluster \cite{antonini13}. While this may not be a classical cluster because its stellar population is not coeval, the `classical' globular clusters seem to be hosting multiple stellar populations too \cite{gratton12}. Unfortunately, cluster formation simulations are presently unable to cover the full dynamic range from the hydrogen burning limit to the relevant galactic scales. A fully self-consistent model for nuclear cluster formation therefore remains out of reach, and analytical or otherwise simplified arguments need to be invoked to estimate the connection between (former) nuclear clusters and globular clusters \cite{kruijssen12b}.

Another mechanism that is important when trying to understand cluster formation in extreme galactic environments is what we recently dubbed the `cruel cradle effect' \cite{kruijssen11,kruijssen12}, i.e.~the disruption of gravitationally bound stellar structure on a cluster formation time-scale due to the tidal perturbations caused by the dense natal environment. The disruption is caused by structures that can be resolved in galaxy-scale numerical simulations \cite{kruijssen11}, and hence can be modelled using sub-grid models. An example is shown in Figure~10, which follows the evolution of the star cluster population in a galaxy merger simulation from \cite{kruijssen11,kruijssen12c}. The simulation models the tidal disruption of the entire cluster population by measuring the tidal field tensor and applying the implied mass loss to each cluster. The merger-induced torques drive up the gas density, which leads to a dramatic increase of the frequency and strength of tidal perturbations. In this simulation, the cruel cradle effect causes the efficient destruction of clusters at $t=0.3$--$0.8$~Gyr, and at $t=1.7$--$2.1$~Gyr.

In \cite{kruijssen12d}, it was shown that the cruel cradle effect only affects the CFE above gas surface densities of $\Sigma\sim10^3~\msun~{\rm pc}^{-2}$, which corresponds to a star formation rate density of a few $\msun~{\rm kpc}^{-2}~{\rm Myr}^{-1}$. Such extreme conditions are reached in galaxy mergers (e.g.~the Antennae \cite{wei12}, Arp~220 \cite{downes98}), starburst galaxies (e.g.~M82 \cite{kennicutt98b}, Haro~11 \cite{adamo11}), high-redshift galaxies \cite{tacconi08,swinbank11}, and galaxy centres \cite{longmore13,kruijssen13}. In the context of long-term cluster disruption rather than the relatively brief window of cluster formation, the enhanced disruption of stellar structure at young ages is important at much lower surface densities \cite{kruijssen11}, as has been confirmed by recent HST observations of M83 \cite{bastian12}.

\subsection{\sc The collective effect of the different physics} \label{sec:coll}
Having discussed the impact of the various physics discussed in \S\ref{sec:physics} on the cluster formation process, it is now possible to add them up. The mechanisms that each individually have been found to directly affect the SFE per free-fall time $\epsilon_{\rm ff}$ are initial turbulence, radiative feedback, and magnetic fields, while the galactic environment provides a second-order factor that may promote cluster formation through other mechanisms (e.g.~nuclear clusters \cite{antonini13} or can cause stellar structure to become unbound on a short time-scale \cite{kruijssen12c}. There exist numerical simulations that combine these different physics. For instance, \cite{krumholz12} have performed large-scale AMR simulations of cluster formation in which initial turbulence, radiative transfer and protostellar outflows are included, and a gaseous environment beyond the star-forming region itself is added to approximate the galactic ISM. Figure~11 shows the stellar masses (and the total number of stars) as a function of time for simulations with different initial conditions and physics. The SFE per free-fall time reached in their most complete simulation is $\epsilon_{\rm ff}\sim0.2$, which is promising, although still much larger than the required $\epsilon_{\rm ff}\sim0.02$. The difference between the blue and red curves illustrates the modest ($\sim20\%$) impact of adding protostellar outflows.
\begin{figure*} \label{fig:krumholz}
\center\includegraphics[width=8cm]{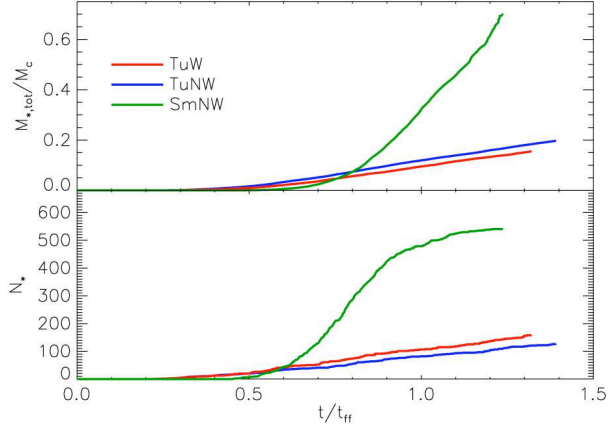}
\caption{Stellar mass (top panel) and total number of star particles (bottom panel) as a function of time for three simulations. The run SmNW is based on smooth, non-turbulent initial conditions and excludes protostellar outflows, and hence it represents pure gravitational collapse. The run TuNW includes initial turbulence, whereas TuW also introduces protostellar outflows. (Figure taken from \cite{krumholz12}, reproduced by permission of the AAS.)}
\end{figure*}

While the combined effect of the different mechanisms on $\epsilon_{\rm ff}$ is approximately additive in the simulations by \cite{krumholz12}, this is generally not the case because the physical processes do not act independently. As holds very often, the grand total is more than just the sum of its parts. An example is given in Figure~12, which shows the result of four AMR simulations \cite{wang10} that include an increasing number of physical mechanisms, all of which influence $\epsilon_{\rm ff}$ quite substantially. Pure gravitational collapse leads to a rapid increase of the stellar mass, and while this is suppressed by turbulence and magnetic fields, the strongest flattening of the mass growth occurs due to the addition of protostellar outflows. Here, they cause $\epsilon_{\rm ff}$ to decrease by a factor of 2--3 rather than the mere 20\% of the previous example, and the SFE per free-fall time becomes $\epsilon_{\rm ff}\sim0.1$ when including magnetic fields, protostellar outflows, and initial turbulence \cite{nakamura07,wang10,gendelev12}. The reason is that magnetic fields link different parts of the cloud, which enhances the efficiency of the deposition of outflow momentum in the ambient ISM. As a result, the protostellar outflows are capable of breaking up the dense filaments that funnel pristine gas to the main sites of star formation -- a sharp contrast with the behaviour in simulations without magnetic fields, in which outflows simply escape along the path of least resistance without affecting the high-density gas. Also interesting to note is that in the presence of magnetic fields, collimated outflows are more effective at changing cloud dynamics (contrary to what is found non-magnetic simulations), because they reach further into the surrounding medium \cite{nakamura07,gendelev12}. This is of crucial importance, because turbulence that is driven on larger scales decays more slowly than smaller-scale driven turbulence, and hence yields a lower $\epsilon_{\rm ff}$. The inclusion of magnetic fields thus does not only affect the impact of outflows, but also the effect of turbulence.

Numerical simulations of cluster formation are capable of {\it approximating} the observed efficiency of cluster formation, provided that the important physics are all included. The simulations by \cite{wang10} did not include radiative feedback, and hence there could be a chance that its interplay with the other relevant mechanisms would provide the final push from $\epsilon_{\rm ff}\sim0.1$ to $\epsilon_{\rm ff}\sim0.02$. This is certainly possible, because (as mentioned in \S\ref{sec:magn}) the influence of magnetic fields largely complements the effect of radiative feedback in terms of their respective spatial ranges \cite{price09}. No simulation has been performed that includes turbulence, magnetic fields, protostellar outflows, and radiative feedback, but \cite{price09} did carry out SPH simulations in which only outflows are missing. They find that while radiative feedback does affect the SFE per free-fall time in the presence of magnetic fields, it only causes a $\sim50\%$ difference rather than the factor of two seen in the simulations by \cite{offner09}. In their run with the strongest magnetic field, \cite{price09} find $\epsilon_{\rm ff}\sim0.1$ when including radiative feedback. Although there is no reason to assume that such an estimate is accurate, a simple addition of the impact of radiative feedback and protostellar outflows suggests that a simulation covering all of the important physical processes would yield SFEs per free-fall time in the range $\epsilon_{\rm ff}=0.04$--$0.07$. While this is clearly a tentative prediction, it is comparable to the value observed for star-forming regions in the solar neighbourhood \cite{evans09}.
\begin{figure*} \label{fig:wang}
\center\includegraphics[width=8cm]{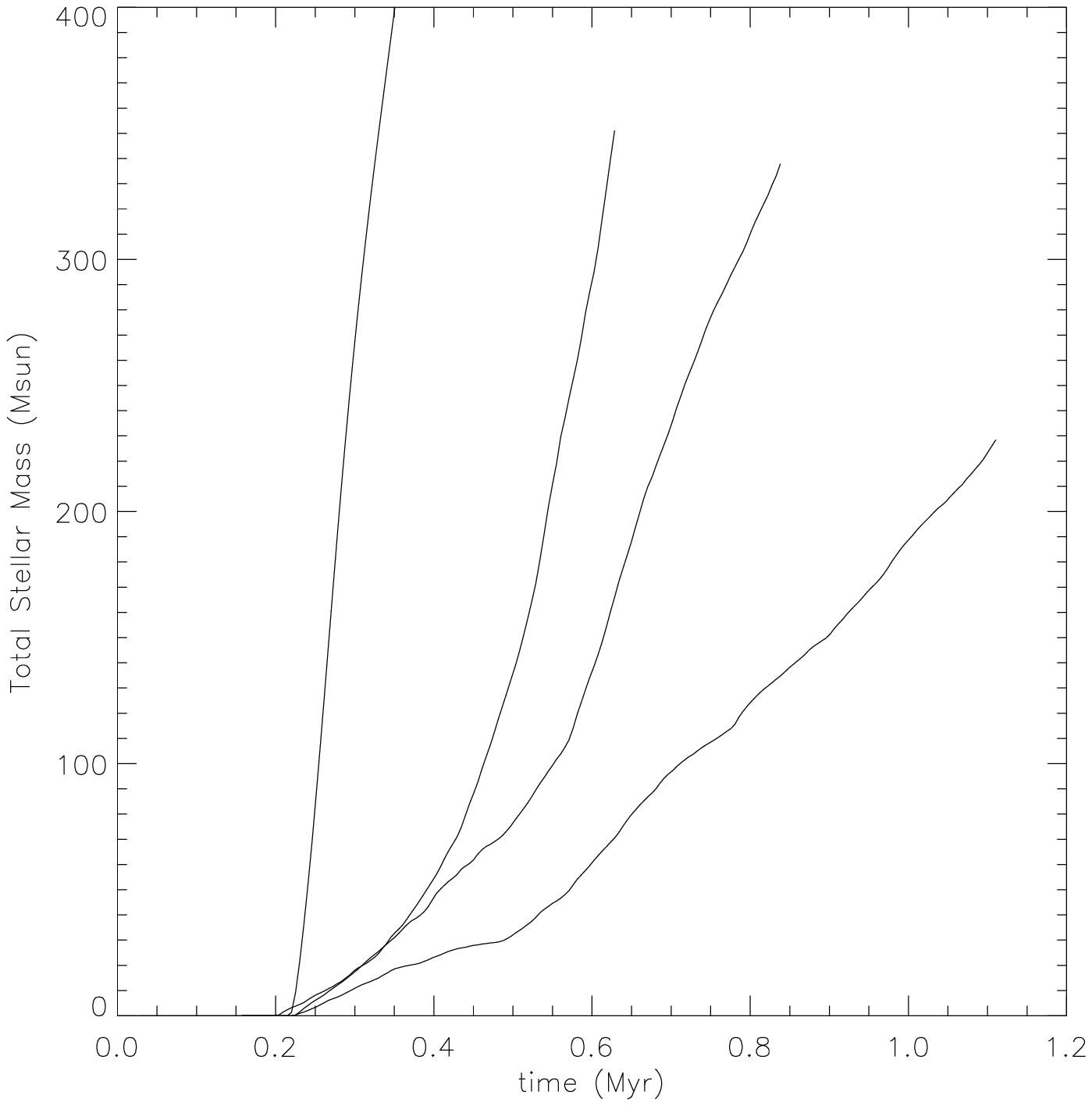}\includegraphics[width=8cm]{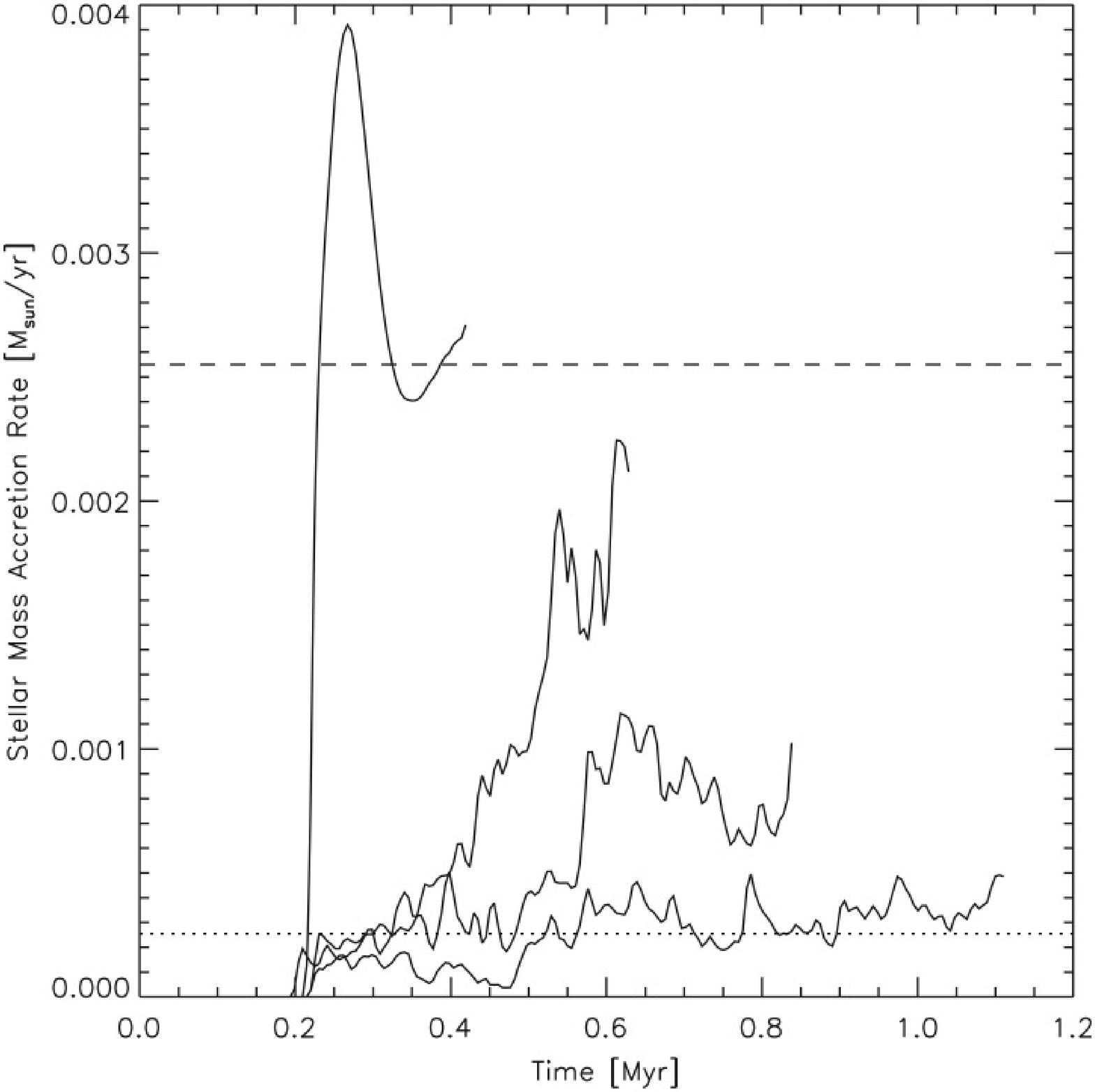}
\caption{{\it Left}: Stellar mass as a function of time in four simulations by \cite{wang10}. The left-most curve indicates a simulation of pure gravitational collapse, and the other curves each add an extra piece of physics to the simulation. From left to right (or top to bottom), they add turbulence, a magnetic field, and protostellar outflows. {\it Right}: Time-evolution of the mass accretion rates of the four simulations in the left-hand panel. The order of the lines is the same as before. The dashed line indicates the characteristic free-fall rate, and the dotted line marks 10\% thereof. (Both panels taken from \cite{wang10}, reproduced by permission of the AAS.)}
\end{figure*}

The current wealth of cluster formation simulations allows the identification of the crucial physics that need to be included to faithfully represent observed star-forming environments. In particular, a turbulent initial velocity field, a magnetic field, protostellar outflows, and radiative feedback seem to be key in slowing down cluster formation to the observed rate. If true, there is indeed a plethora of mechanisms that gravity needs to overcome in order to form a gravitationally bound cluster. This answers the main question posed in \S\ref{sec:role}: how to {\it not} form a cluster? At low densities, the discussed physics are capable of suppressing the SFE and keeping the stars in an unbound configuration, whereas at higher densities (1) the free-fall time is too short for this to occur, and (2) the efficiency of the involved feedback mechanisms (outflows, radiation) is suppressed by the short dissipation time-scale. The transition between both density regimes should exhibit some environmental variation, but for solar neighbourhood conditions it is expected occur at a density of $\rho\sim10^3~\msun~{\rm pc}^{-3}$ (see Figure~1 of \cite{kruijssen12d}). Cluster formation simulations often model regions of very high density and the above number shows that this is reasonable -- clusters form at the high-density end of the ISM density spectrum. Whether or not this accurately reflects the formation environment of most stars is a very different question, and the preliminary answer seems to be negative -- only a minority of all stars are born in clusters \cite{lada03,goddard10,kruijssen12d}.

\section{\sc Discussion and open questions} \label{sec:disc}
Even though numerical simulations of star cluster formation yield a reasonable match to observed star-forming regions, this does not necessarily mean that we have a complete picture of the physics that govern cluster formation. The aim of running numerical simulations is to simplify nature to such a degree that the physics can be understood. However, numerical modelling often depends on certain tricks or compromises that could lead to spurious results. In that respect, the broad agreement between SPH and AMR simulations is reassuring. These are widely different numerical approaches, and it is very unlikely that they are wrong in the same way.

In the following, I summarize some of the directions in which further work is needed, and I will pay particular attention to those questions that were considered during the discussion session at the `Formation and Early Evolution of Stellar Clusters' workshop where this review was presented. In a select number of cases, and whenever appropriate, I have taken the liberty of including the thoughts of others, but unfortunately without having a record of the individuals responsible for the particular contributions.

From the theoretical side, a clear picture needs to be obtained of which physical mechanisms are essential for modelling cluster formation. While this review has presented a brief overview of which physics are likely important, one of its main conclusions is also that no entirely satisfactory cluster formation simulation has been run. Hence, we do not know what the details are of the final interplay between the several processes, and mapping their combined influence in more detail will be one of the main challenges for the upcoming years.

The other part of the story is that the simplified nature of numerical simulations allows one to isolate certain physics and study their effect in detail -- adding more physics is not always better. One of the main recent discussion points has been to understand how the gas and stars are distributed with respect to each other, which has been motivated by the discovery that young clusters in numerical simulations are embedded in 2D, but not in 3D (see \S\ref{sec:bound} and \cite{kruijssen12}). This is of crucial importance -- if the gas merely forms a cocoon of gas around some virialized stellar distribution, it will not have any effect on the gravitational boundedness and hence cluster formation is unavoidable. Modelling cluster formation is therefore impossible without following the gas hydrodynamics. Without a self-consistently modelled gas component, the response of the stars to gas removal follows trivially from the initial conditions. Pure $N$-body simulations that include the gas as a background potential therefore need to adopt initial conditions that are motivated by simulations in which the dynamical interaction between gas and stars is accounted for. An improved, generalized picture of how this interaction proceeds would be a very important asset to the field, and does not necessarily rely on including all of the physical mechanisms discussed in \S\ref{sec:physics}.

Furthermore, and at the risk of sounding obvious, detailed studies of simplified simulations enable the techniques used in current cluster formation modelling to be improved. Compromises are being made in wind, outflow, and radiative transfer models, and even in the hydrodynamics itself. Developments such as new SPH and sink particle algorithms \cite{hopkins13,hubber13} are excellent examples of how simulations that do not include all of the physics relevant on large scales are crucial for advancing our understanding of cluster formation.

In the long run, there will be a persistent interest in scaling up the simulations to start probing the formation of YMCs or perhaps even globular clusters. Such mass scales can presently only be reached by increasing the sink particle mass, thereby foregoing the direct correspondence to the stellar mass spectrum, and consequently omitting the reasonable modelling of the stellar dynamics. A bigger computer may not be the only solution to this apparent problem -- in terms of its global properties, a cluster formation simulation is not sensitive to the stellar mass spectrum or stellar dynamics while it is younger than a relaxation time. On the mass scales of YMCs and globular clusters, this time window should be sufficiently long to cover a non-negligible part of cluster formation. It also allows access to a broader range of questions, such as explaining the appearance of multiple generations of stars in massive clusters \cite{gratton12} or the possible connection between cluster formation and the origin of massive black holes \cite{portegieszwart04,luetzgendorf13,kruijssen13b}.

Despite all the uncertainties and potential for improvement, cluster formation simulations look surprisingly similar to observed star-forming regions. This by itself already warrants a comparison between theory and observations, but it is not obvious how this comparison should be made. How far should theorists go in trying to mimic observational data? Clearly, the comparison benefits from a situation in which numerical and observational work meet in the middle. There has been substantial progress in converting simulated data to observables \cite{ercolano12}, but even these state-of-the-art models show what a simulation would look like using a perfect instrument, and if it were the only object in the Universe. Close collaboration between observers and modellers is necessary to overcome these and other disconnects.

Observational work can also play a key role in mapping the relative behaviour of stars and gas. As mentioned several times in this review, it is essential for theory to know where the gas and stars are with respect to each other. A comparison of observational data with the recent numerical results would be very helpful, both for understanding the relevant physics, and for driving the development of improved initial conditions for early cluster evolution modelling. With high-resolution interferometers such as ALMA and the EVLA, it is possible to derive the detailed phase-space structure of the gas, whereas Gaia will provide such information of the stars. The observations obtained with these facilities will be instrumental in answering this question.

Perhaps the main observational stimulus for theoretical and numerical progress is the discovery of freaks -- those objects that challenge existing models and allow us to probe the extremes of parameter space. The recent rediscovery of the densest known Galactic GMC G0.253+0.016 (`the Brick', \cite{longmore12}) is an excellent example. Surprisingly, it shows no signs of star formation, and with the upcoming availability of high-resolution ALMA observations, such clouds will likely teach us much more about star and cluster formation than objects that we understand. It has also been suggested that the Brick is the youngest of a series of clouds that follow an evolutionary sequence \cite{longmore13b}. If true, this opens up the exciting avenue that snapshot sequences from numerical simulations can be verified against a real time-series existing in nature.

From the numerical, theoretical, and observational directions, the gaps between ISM and stellar-dynamical astrophysics, and Galactic and extra-galactic cluster formation studies are gradually being bridged. Cluster formation simulations have matured, and the coming years will see a wide range of new questions that need answering. This should allow the field of cluster formation simulations to maintain its rapid development of the past decade into the next -- hopefully this review will be outdated very soon.

\section*{\sc Acknowledgements}
I am grateful to the organizers of the Sexten Cluster Workshop -- Nate Bastian, Rob Gutermuth and Linda Smith -- for organizing a very pleasant meeting and for inviting me to give a review talk on this exciting topic. I would particularly like to thank my colleagues on the theory panel -- Cathie Clarke, Jim Dale and Simon Goodwin -- for instigating a very fruitful plenary discussion and for allowing me to use their ideas and remarks in \S\ref{sec:disc}. Finally, I thank Volker Springel, Bruce Fryxell, Philipp Girichidis, Charles Hansen, Jim Dale, Stella Offner, Inti Pelupessy, Paolo Padoan, Mark Krumholz, and Zhi-Yun Li for giving me permission to reuse their figures in this review.

\bibliography{mybib}

\end{document}